\documentclass{aastex701}

\usepackage{xspace}
\usepackage{hyperref}
\usepackage{color}
\usepackage{amsmath}

\newcommand{\JWST}{{\it JWST}\xspace}

\newcommand{\Teff}{\ensuremath{T_{\mathrm{eff}}}\xspace}
\newcommand{\logg}{log~$g$\xspace}
\newcommand{\Msol}{\ensuremath{M_{\odot}}\xspace}
\newcommand{\Rsol}{\ensuremath{R_{\odot}}\xspace}
\newcommand{\fPBH}{\ensuremath{f_{\rm PBH}}\xspace}

\newcommand{\frametime}{21.474~s\xspace}

\begin{document}

\title{Fast Lensing And Sub-minute High-accuracy photometry (FLASH) with JWST:\\Pipeline and Primordial Black Hole Microlensing Constraints from M31}

\newcommand{\uhawaii}{\affiliation{Department of Physics and Astronomy, University of Hawai`i at M{\=a}noa, Honolulu, Hawai`i 96822}}
\newcommand{\lbnl}{\affiliation{Physics Division, E.O. Lawrence Berkeley National Laboratory, 1 Cyclotron Rd., Berkeley, CA, 94720, USA}}

\author[0000-0001-5402-4647]{David Rubin}
\email{drubin@hawaii.edu}
\uhawaii
\lbnl

\author[orcid=0000-0003-2274-0301]{Istv\'an Szapudi}
\affiliation{Institute for Astronomy, University of Hawai`i at M{\=a}noa, Honolulu, Hawai`i 96822}
\email{istvan@hawaii.edu}

\begin{abstract}

Minute- and sub-minute microlensing provides a direct probe of compact dark matter at asteroid-mass scales, where primordial black holes with $M_{\rm PBH}\lesssim 10^{-10}M_\odot$ would produce short-duration, finite-source-suppressed amplifications of stars in extragalactic fields. To enable these measurements, we present \textsc{FLASH}: Fast Lensing And Sub-minute High-accuracy photometry with JWST, a pipeline designed to recover such events from JWST/NIRCam imaging by preserving detector-ramp time information and performing high-precision crowded-field photometry. \textsc{FLASH} begins from Level~1 NIRCam data products, applies microlensing-oriented detector calibration (including improved treatment of non-linearity and position-dependent detector calibration), and extracts forced multi-band light curves in crowded fields. We apply the pipeline to two public JWST/NIRCam datasets targeting the disk of M31 with 21.5~s cadence: GO-4735 and GO-2609.  Although these observations are suboptimal as a microlensing survey, they provide a stringent test of JWST time-domain performance in a high-surface-brightness, crowded stellar field.  
Using these data with our own microlensing simulations, we derive the first JWST-based microlensing constraints toward M31, finding $\fPBH \lesssim 100$ for $M_{\mathrm{PBH}} \sim 10^{-9} \Msol$. Our results demonstrate that JWST is a competitive platform for high-cadence microlensing and that a dedicated, continuous NIRCam monitoring program could substantially improve constraints on compact dark matter below the current ground-based frontier.

\end{abstract}

\keywords{James Webb Space Telescope, Dark matter, Gravitational nanolensing, Gravitational microlensing, Astronomical detectors}

\section{Introduction}

The nature of dark matter remains one of the central unsolved problems in astrophysics, despite nearly a century of evidence for gravitational mass not accounted for by the observed baryonic component \citep{Zwicky1933,RubinFord1970,Clowe2006,Planck2020}.  The possible explanations span a wide theoretical landscape: new weakly interacting particles motivated by extensions of the Standard Model, ultralight fields, modifications of gravity, and compact objects formed in the early Universe.  Among the latter, primordial black holes \citep[PBHs; e.g.,][]{Hawking1974BlackHoleExplosions,Carr1975PBHMassSpectrum} occupy a distinctive position.  They require no new stable particle species, interact gravitationally by construction, and, in principle, can be produced over a broad range of masses via enhanced primordial density fluctuations, phase transitions, cosmic strings, or other early-Universe mechanisms.  At the same time, decades of direct, indirect, and astrophysical searches have not produced a conclusive non-gravitational detection of particle dark matter \citep[e.g.,][]{Jungman1996SupersymmetricDM,Hooper2018TASIIndirect,Arina2018SimplifiedModels, LZ2025}.  The question therefore remains open: can any allowed population of compact objects account for a significant fraction of the dark matter?

PBHs are especially compelling since the relevant phenomenology is almost entirely gravitational.  The discovery of merging stellar-mass black holes by LIGO/Virgo \citep{LIGO2016GW150914} and horizon-scale imaging of supermassive black holes \citep{EHT2019M87I} demonstrate that black holes are abundant astrophysical objects, while several authors have explored whether some gravitational-wave events or high-redshift massive black holes could point to a primordial component \citep[e.g.,][]{Sasaki2016GW150914PBH,Bird2016DidLIGODetectDM, Zhang2025}.  These interpretations remain debated, and stellar evolution certainly produces black holes in some mass ranges.  For the dark-matter problem, however, the key issue is not whether black holes exist, but whether a cosmologically significant abundance of PBHs can survive the many existing constraints \citep{Carr2021,Green2024,Gorton2024}.  This is particularly important because a realistic PBH mass function need not be monochromatic.  Extended distributions, including lognormal or power-law forms with cutoffs, can shift the interpretation of single-mass limits and leave discovery space near the edges of existing constraints \citep{Ballesteros2018}.  Each improvement at the low-mass frontier, therefore, tests both a new mass scale and a new class of extended mass functions.

Gravitational microlensing is one of the most direct ways to search for PBHs \citep{PressGunn1973,Paczynski1986Microlensing,Griest1991,Alcock1996MACHO,Alcock1997MACHO,Tisserand2007EROS2,Griest2011KeplerMicrolensing,Mroz2017Nature548183,Niikura2019bPhysRevD}.  When a compact lens passes close to the line of sight to a background star, the apparent stellar flux is temporarily amplified with a characteristic Paczyński light-curve shape.  For a lens of mass $M$ at fractional distance $x=D_l/D_s$ to a source at distance $D_s$, the physical Einstein radius is
\begin{equation}
R_E = \left[\frac{4GM}{c^2}D_s x(1-x)\right]^{1/2},
\end{equation}
and the corresponding Einstein crossing time is $t_E=R_E/v_\perp$.  Toward M31, adopting nominal values $D_s= 760~{\rm kpc}$ \citep{Li2021}, $x(1-x)=0.01$, and $v_\perp=200~{\rm km~s^{-1}}$ gives
\begin{equation} \label{eq:tE}
t_E \simeq 1~{\rm min}
\left(\frac{M}{10^{-10}M_\odot}\right)^{1/2}
\left[\frac{x(1-x)}{0.01}\right]^{1/2}
\left(\frac{v_\perp}{200~{\rm km~s^{-1}}}\right)^{-1}.
\end{equation}
(For these parameters, the Einstein radius is in the nanoarcsecond range as viewed from Earth, so this is actually nanolensing.) Thus, the PBH mass range near and below $10^{-10}M_\odot$ naturally produces minute-scale, and eventually sub-minute-scale, Einstein crossing times.  At sufficiently low masses, however, the event duration no longer scales only with $R_E$.  Finite-source effects become important when the Einstein radius projected into the source plane is comparable to the stellar radius, or equivalently when $R_E \lesssim xR_\star$ in the lens plane.  In this regime, the maximum amplification is suppressed, and the characteristic duration is set by the time required for the lens to cross the projected stellar disk,
\begin{equation}
t_\star \equiv \frac{xR_\star}{v_\perp}
\simeq 35~{\rm s}
\left(\frac{x}{0.01}\right)
\left(\frac{R_\star}{R_\odot}\right)
\left(\frac{v_\perp}{200~{\rm km~s^{-1}}}\right)^{-1},
\end{equation}
with a full source-diameter crossing lasting $\sim 2t_\star$.  For giant stars in M31, this finite-source timescale can be several minutes or longer even when $t_E$ is formally shorter.  Consequently, the lowest-mass events accessible to JWST are not arbitrarily brief: their amplitudes decrease as the source becomes larger than the Einstein ring, while their observable durations approach the stellar-radius crossing time.  Sensitivity to this regime therefore requires not only a large number of monitored stars, but also high cadence, stable photometry, and accurate knowledge of stellar radii and luminosities.

M31 has long been recognized as a powerful microlensing target \citep{Crotts1992,Baillon1993,Gould1996PixelLensing,CalchiNovati2005POINTAGAPE,Niikura2019a,Smyth2020, Sugiyama2020,Sugiyama2026HSCM31PBH}.  Its stellar disk and bulge provide an enormous source population within a compact area on the sky, while the line of sight probes both the Milky Way and M31 halos.  The high surface brightness that makes M31 attractive also makes it difficult: ground-based observations are limited by atmospheric seeing, unresolved blends, variable backgrounds, and the need to separate microlensing events from intrinsic stellar variability, novae, detector artifacts, and cosmic rays.  These difficulties are especially severe for the shortest events, where only a few data points may sample the light curve.

JWST/NIRCam is therefore a promising facility for low-mass PBH microlensing.  JWST operates at L2 in a thermally stable environment and delivers diffraction-limited near-infrared imaging with substantially reduced crowding relative to wide-field ground-based surveys \citep{Gardner2023JWSTMission,Rigby2023JWSTPerformance,Rieke2023NIRCamPerformance}.  NIRCam observes simultaneously in short- and long-wavelength channels, enabling contemporaneous color information that can help reject many classes of astrophysical and instrumental false positives.  In full-frame imaging, the detector frame time can be as short as $10.737~{\rm s}$, and thus the non-destructive readout ramps contain time-domain information on timescales far shorter than a conventional ramp-fit (Level~2) product.  These properties are well matched to microlensing by PBHs with $M\lesssim 10^{-10}M_\odot$, for which the relevant timescales are minutes or less.  They motivate a reduction strategy that treats JWST imaging not only as a source of deep static images, but as a high-cadence photometric time series embedded within the detector ramps.

 Unfortunately, JWST data are not automatically optimized for this experiment.  The standard JWST calibration pipeline is designed to produce high-quality calibrated images and spectra for a broad range of observing modes \citep{Bushouse2024JWSTPipeline}.  However, a sub-minute microlensing search imposes different requirements.

In this paper, we introduce \textsc{FLASH}: Fast Lensing And Sub-minute High-accuracy photometry with JWST.\footnote{\url{https://github.com/rubind/NIRCam_FLASH}} \textsc{FLASH} is a JWST/NIRCam microlensing pipeline designed to search for short-duration lensing events in crowded fields using JWST imaging data. The pipeline begins with Level~1 uncalibrated data (and uses Level~3 mosaics for source detection), it applies ramp-level calibrations, applies and improves flat fielding and linearity treatment for differential time-domain photometry, and extracts forced multi-band light curves for large numbers of sources.

This paper applies \textsc{FLASH} to two public JWST/NIRCam datasets targeting the disk of M31.  
  These programs were not designed as microlensing surveys: the observations are dithered, relatively short, and optimized for NIRSpec target selection and stellar-population science rather than uninterrupted time-domain monitoring.  This makes them a conservative but valuable first test of the method.  Any constraints derived from these data therefore demonstrate the capabilities of JWST and \textsc{FLASH} and guide the design of future dedicated JWST microlensing programs.

The structure of the paper is as follows.  Section~\ref{sec:data} describes the JWST/NIRCam observations and ancillary stellar-population information and presents the \textsc{FLASH} calibration, astrometry, and photometry pipeline. Section~\ref{sec:candidates} describes the transient search, the candidate-vetting procedure, and our candidates.  Section~\ref{sec:PBHConstraints} presents our Monte-Carlo simulations and microlensing detection (including using PHAT HST observations of each target star to estimate its individual radius), and converts the search results into PBH constraints. Section~\ref{sec:conclusions} summarizes and discusses the implications for future JWST microlensing programs. The Appendices contain details of our pipelines and simulations that would break the flow of the main paper. Appendix~\ref{sec:wave} discusses the (small) impact of wave optics on our expected light curves; Appendix~\ref{app:event_rate_mc} discusses details of our Monte-Carlo calculations.

\section{Data Selection and Processing} \label{sec:data}
\label{sec:data}

\subsection{Data Selection}

\begin{figure*}[htbp]
\centering
    \includegraphics[width=0.9\textwidth]{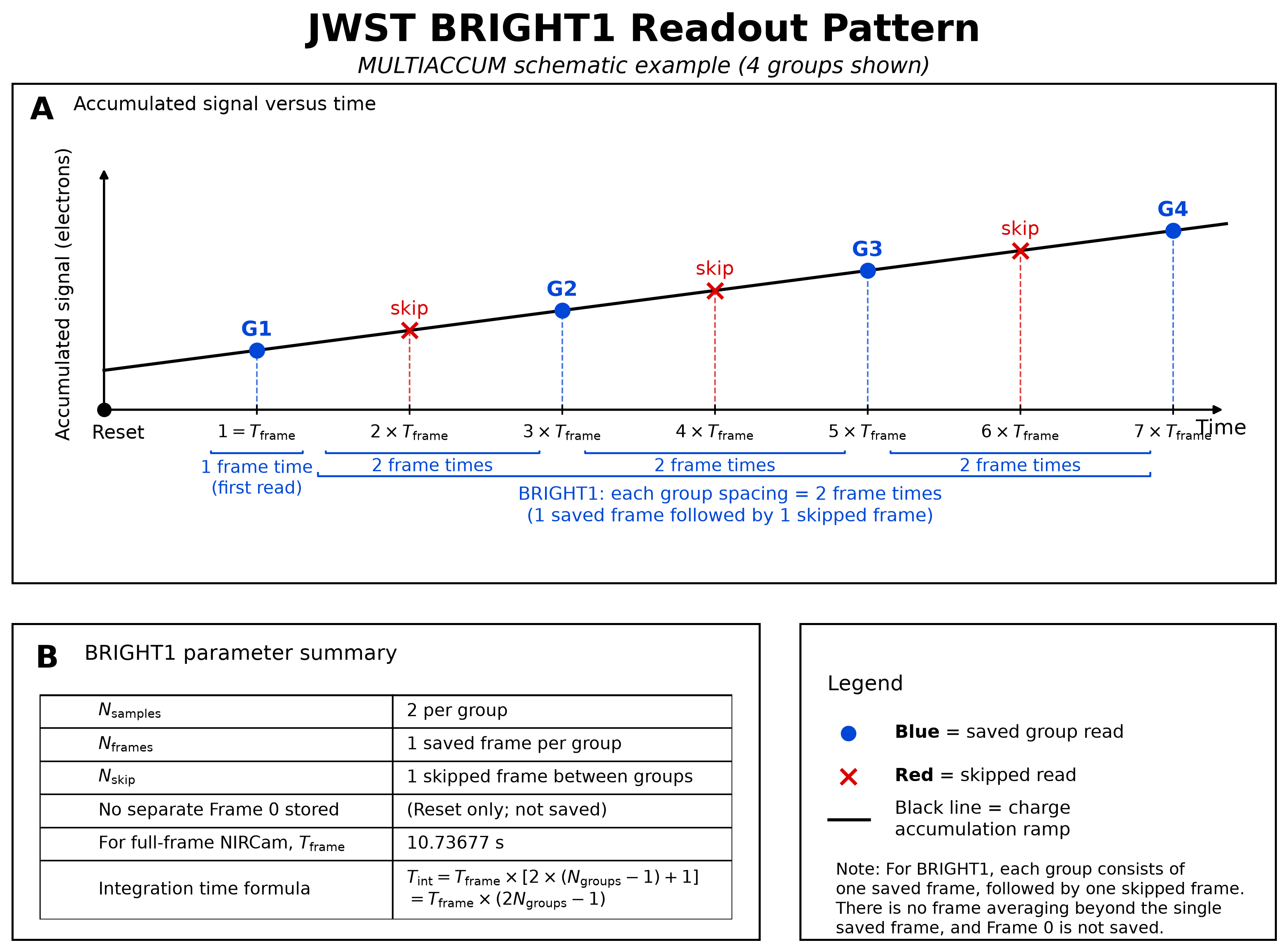}
    \caption{Diagram of the BRIGHT1 readout pattern used in the data we examine. \label{fig:readout}}
\end{figure*}

To find data to develop and test our pipeline, we searched the MAST \JWST archive\footnote{\url{https://mast.stsci.edu/search/ui/\#/jwst}} for programs with the following characteristics:

\begin{itemize}
    \item A target outside the Milky Way but in the Local Group, enabling a long-distance baseline with individually detected stars
    \item A high stellar density, as thousands of star-hours are needed for a competitive microlensing constraint
    \item A significant amount of NIRCam imaging exposure time ($\sim$~hours) in broadband filters
    \item A readout pattern with a fast cadence, ideally RAPID (10.737~s per frame) or BRIGHT1 (21.474~s per frame, see Figure~\ref{fig:readout}) where no grouping of readouts is done so that cosmic rays only affect one consecutive group difference
\end{itemize}

We decided that two NIRCam datasets would be useful for a good first attempt. The first is GO-4735, ``A Closer Look at the Formation and Evolution of M31's Inner Disk'' (PI: Nathan Sandford), which includes NIRCam pre-imaging of two M31 fields in F150W/F277W.  The observations used here include sequences of 16 dithers with seven groups per exposure in the BRIGHT1 readout pattern, corresponding to $2233.248~{\rm s}$ of elapsed observing time per field.  The second is GO-2609, ``Searching for the Alpha-Abundance Bimodality in the M31 Disk'' (PI: David Nidever), which includes NIRCam pre-imaging of two M31 disk locations in F150W/F277W, with eight dithers, five groups, and two integrations, corresponding to $1631.989~{\rm s}$ per field.

These data are not ideal for our purposes; the dithers are quite large, so we must flat-field (and take flat-fielding uncertainties). The very broad $F150W2/F322W2$ filters would have provided much better sensitivity for most of the stars. Nevertheless, this enables a proof of concept, as we discuss below.

\subsection{Data Processing}

We downloaded the calibrated \texttt{\_cal.fits} files for selecting stars and the uncalibrated \texttt{\_uncal.fits} files for performing our rapid-cadence photometry. We processed these into photometric measurements as follows.

\subsubsection{Star Selection}

To select stars, we used \texttt{Tweakreg} to align all \texttt{\_cal.fits} files to Gaia DR3 \citep{GaiaCollaboration2023}. Then we used \texttt{ResampleStep} to resample and combine the $F150W$ data into one mosaic. Because $F150W$ provided the highest S/N, we used it as the basis for our star selection.

We selected sources greater than 30$\times$ the NMAD of the mosaic image, then cut them out in 15$\times$15 pixel cutouts, scaled the cutouts to the same peak, and took the median to estimate the mosaic-image PSF. We correlated this PSF with the mosaic image and selected peaks with a correlation coefficient $> 0.75$. This technique effectively selected stars while avoiding diffraction spikes. In the end, we selected 240,048 stars out of GO-4735 
and 224,156 stars out of GO-2609. 

\subsubsection{Calibrating the \texttt{\_uncal.fits} Images}

For each \texttt{\_uncal.fits} image, we ran: \texttt{GroupScaleStep}, \texttt{DQInitStep}, \texttt{SaturationStep}, \texttt{SuperBiasStep}, \texttt{RefPixStep}, \texttt{LinearityStep}, and \texttt{DarkCurrentStep}. This is a similar set of processing to \url{https://github.com/exonik/jwst_pipeline_testing/blob/master/Stage%201%20testing%20-%20NIRCam.ipynb}. However, we exclude the jump detection as cosmic rays and microlensing events can both cause rapid increases in flux, and we do not want to exclude the latter. We also exclude the ramp fitting. We then correct each image by dividing by the flat field and the pixel area map. We also convert each image to electrons, which simplifies the Poisson-noise calculation and matches the units of the read-noise.

\subsubsection{\texttt{\_uncal.fits} PSF Derivation}

We built one pixel-convolved PSF (effective PSF or ePSF) per detector and fit the photometric impact of any ePSF variation over the detector with ubercalibration terms described in Section~\ref{sec:starflats}.

We processed each \texttt{\_uncal.fits} image by converting each ramp to a series of frame-to-frame differences and masking any difference that includes a saturated frame. We built a median of all frame differences, constructed a smooth sky background with \texttt{Background2D} and \texttt{SExtractorBackground}, and subtracted this from the sets of differences.

We converted the star catalog to $x$ and $y$ locations for each image using the \texttt{Tweakreg}-aligned \texttt{\_cal.fits} images which have the same pixel coordinates as the \texttt{\_uncal.fits} images. We examined stars at least 15 pixels from all other stars and removed any stars at the image edges. Then, we extracted 21$\times$21 pixel cutouts centered on each surviving star. We built a rough median ePSF from the cutouts, then fit each cutout to obtain an approximate flux. Then, we examined each star's residuals, computing 1.4826 times the median absolute deviation of the residuals (NMAD, a robust measure of the dispersion) as a fraction of the star's flux. We rejected all stars where the NMAD of the residuals (compared to the flux of the star) is more than 5$\times$NMAD from the median of all the NMADs. After this structured-background cut, we dropped all but the brightest 200 stars. We then iterated, re-estimating the ePSF, estimating the star fluxes, evaluating the residuals again and dropping any stars with high background dispersions. Finally, we kept the brightest 100 stars in each \texttt{\_uncal.fits} image to build the ePSF.

\newcommand{\PSFsentence}{Our ePSFs have high signal-to-noise and no obvious artifacts.\xspace}

\begin{figure*}[htbp]
\centering
    \includegraphics[width=0.9\textwidth]{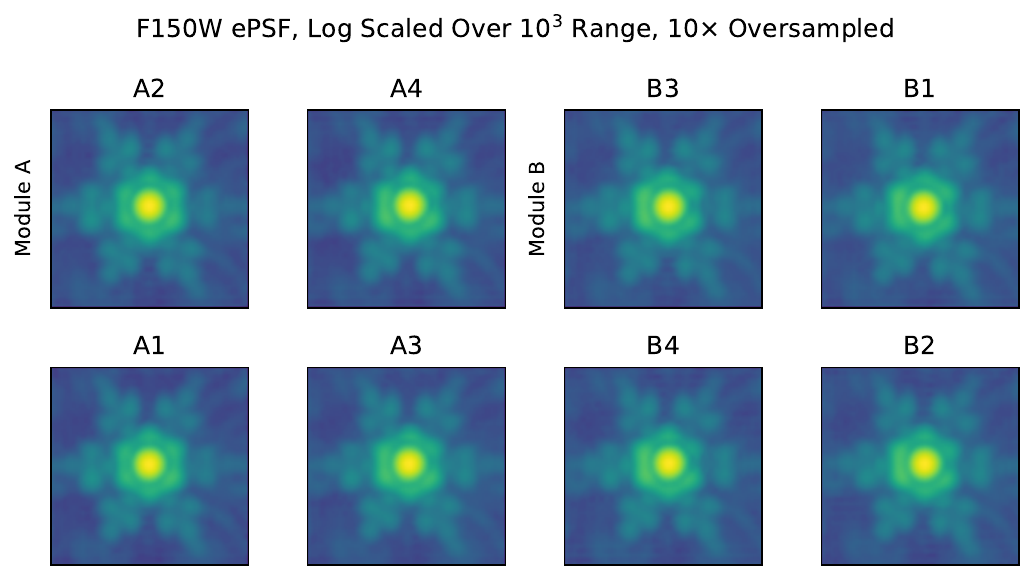}

    \includegraphics[width=0.45\textwidth]{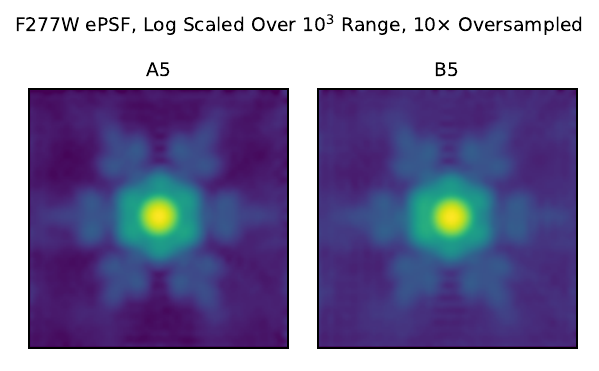}
    \caption{ePSFs for the GO-4735 data shown on a logarithmic scale for three orders of magnitude down from peak. \PSFsentence \label{fig:ePSF}}
\end{figure*}

We wrote a custom ePSF-building code that alternates between estimating each star's centroid, flux, and (spatially constant) sky background, and estimating the ePSF using all stars. We assumed and enforced uniform sub-pixel response (a uniform square pixel). For robustness, we minimized the absolute sum of residuals (optimal for a fat-tailed Laplace distribution for the residuals), rather than the more conventional sum of squared residuals (optimal for Gaussian-distributed residuals). We modeled the PSF with a 21$\times$21-pixel 2D spline with $2\times$ oversampling (two nodes by two nodes for each pixel, so 42$\times$42 spline nodes). After interpolating this spline to $10\times$ oversampling, we convolved it by a uniform pixel, forming a $10\times$ oversampled ePSF. We then fit this ePSF model to the data. Figure~\ref{fig:ePSF} shows the ePSFs constructed from GO-4735 data; the ePSFs constructed from the GO-2609 data are visually similar. \PSFsentence

\subsubsection{\texttt{\_uncal.fits} PSF Photometry} \label{sec:PSFPhotometry}

With the ePSFs in hand, we performed PSF photometry using the differences between consecutive readouts in our calibrated \texttt{\_uncal.fits} files. We took the median of all differences for a given exposure, computed the centroid, and then performed forced photometry at that location for all differences. Forced photometry is more robust to low per-frame-difference signal-to-noise and is much more robust to cosmic-ray hits. In addition to estimating the ePSF amplitudes, we simultaneously modeled the sky background by taking the median (data $-$ scaled ePSF).

We also computed the ePSF-weighted RMS residual for all measurements. This RMS was typically a few percent for good measurements but became much larger if a cosmic-ray hit significantly overlapped the star. We dropped measurements with RMS residuals above 20\% of the peak flux.

Any star measurement with more than 50\% of its flux on flagged pixels was rejected; only a few pixels contain half the flux, and a saturated pixel also flags the neighboring $3\times3$ pixels, so the saturation of just the peak of the star caused the measurement to be dropped.

\subsubsection{Photometric-Catalog-Level Linearity Refinement}

The NIRCam linearity correction is based on a polynomial mapping between observed counts and corrected counts \citep{Canipe2017}. This procedure leaves systematic residual wiggles that mostly average out for full ramps (e.g., \citealt{Canipe2017} Figure~3). But these wiggles do not average out for parts of ramps, which is what we are photometering in this work. Computing a more flexible linearity correction would likely remove these systematic residuals, but the linearity data are not public. Thus, we compute a catalog-based linearity correction based on our data.

For each detector, filter, and ramp frame, we apply an empirical spline correction as a function of observed flux to align the flux to the ramp median. We include only good photometric measurements (as above, ePSF-weighted RMS residuals below 0.2 of peak, and $> 50\%$ of flux on good pixels). We use a cubic spline with 11 spline nodes uniformly spaced in flux from 1000 $e^-$ to the maximum observed flux. For robustness, we minimized the absolute sum of pulls ($\equiv$ residual/uncertainty), essentially assuming a Laplace distribution for the pulls.

This linearity correction also includes apparent nonlinearity from the impact of the brighter-fatter effect \citep{Antilogus2014, Hirata2020} on the photometry, so the linearity correction could depend on sub-pixel position. However, we have looked for a sub-pixel dependence, and it is small, even in the undersampled F090W (data not considered in this work), so we ignore it for now.

\subsubsection{Flat-Field Refinement Derived with Ubercalibration} \label{sec:starflats}

\newcommand{\flatsentence}{We obtain similar looking corrections for GO-2609, even though the data are independent and taken at a different time.\xspace}

\begin{figure*}
\centering
    \includegraphics[width=0.9\textwidth]{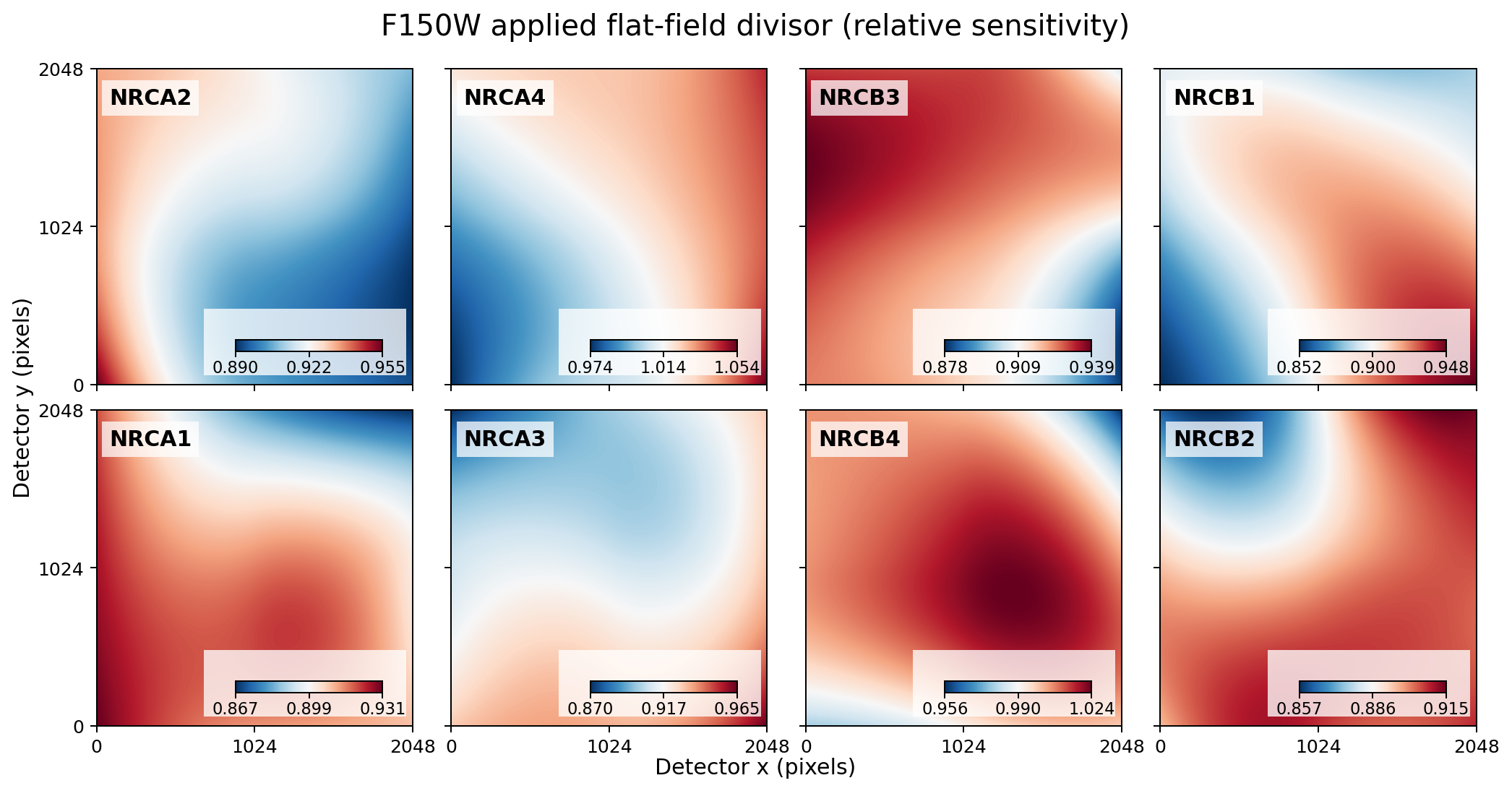}

    \includegraphics[width=0.5\textwidth]{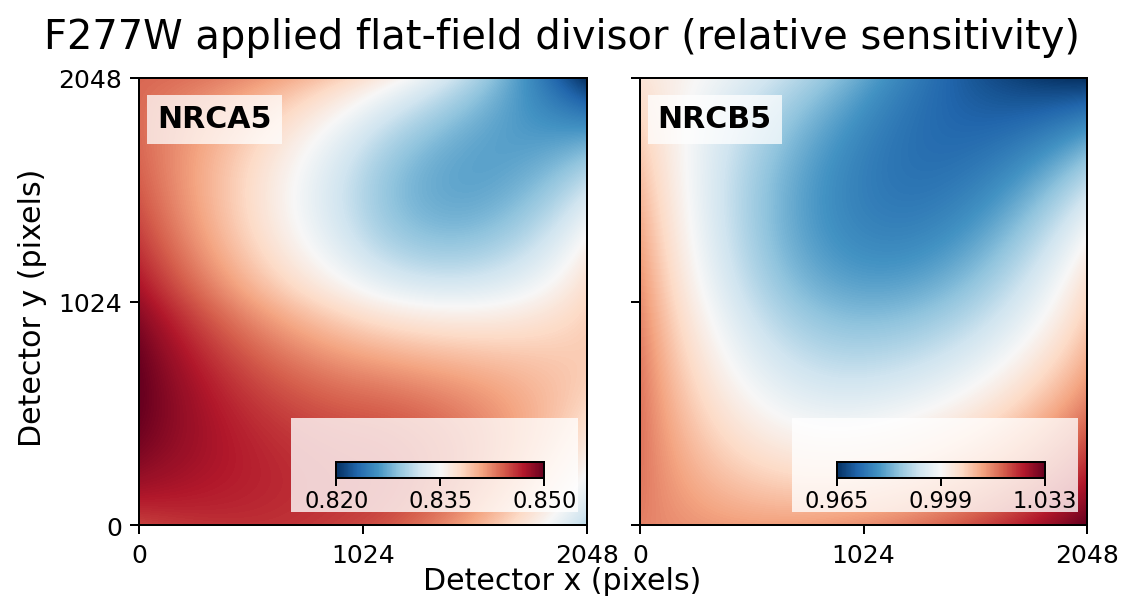}

    \caption{Applied flat field for GO-4735; the corrections are several hundredths of a magnitude over each detector (note that we work in units of electrons, so some of the detector-to-detector variation is due to different gains). It is thus critical that we apply these corrections to our well-dithered photometry. \flatsentence These filter-dependent corrections are so large that they may have implications for distance-ladder work with JWST (e.g., \citealt{Riess2024, Freedman2025}). \label{fig:flatfield}}
\end{figure*}

Examining our light curves at this point showed that dithering frequently produced significant jumps in flux, indicating that we needed to refine the flat field. We thus computed an empirical, filter-specific sensitivity correction from our dithered stellar photometry. For each exposure, we rejected measurements with PSF-weighted residual RMS above 0.2, required at least four valid measurements, and used the median flux. For stars brighter than 10,000 $e^-$, we then fitted jointly each star's unknown intrinsic flux and each detector's relative normalization with a two-dimensional spline grid ($4\times4$ nodes over each detector). One detector normalization is fixed to unity in one spot to remove the degeneracy between stellar flux and overall sensitivity.
As with the PSF fit, we minimized the absolute sum of residuals, making the fit less sensitive to outliers than ordinary squared residuals. Figure~\ref{fig:flatfield} shows the derived correction for GO-4735. \flatsentence

With the linearity and flat-field refined, we had a science-ready set of photometric measurements.

\subsection{FLASH Photometric Performance}

\begin{figure*}
    \centering
    \includegraphics[width=0.95\textwidth]{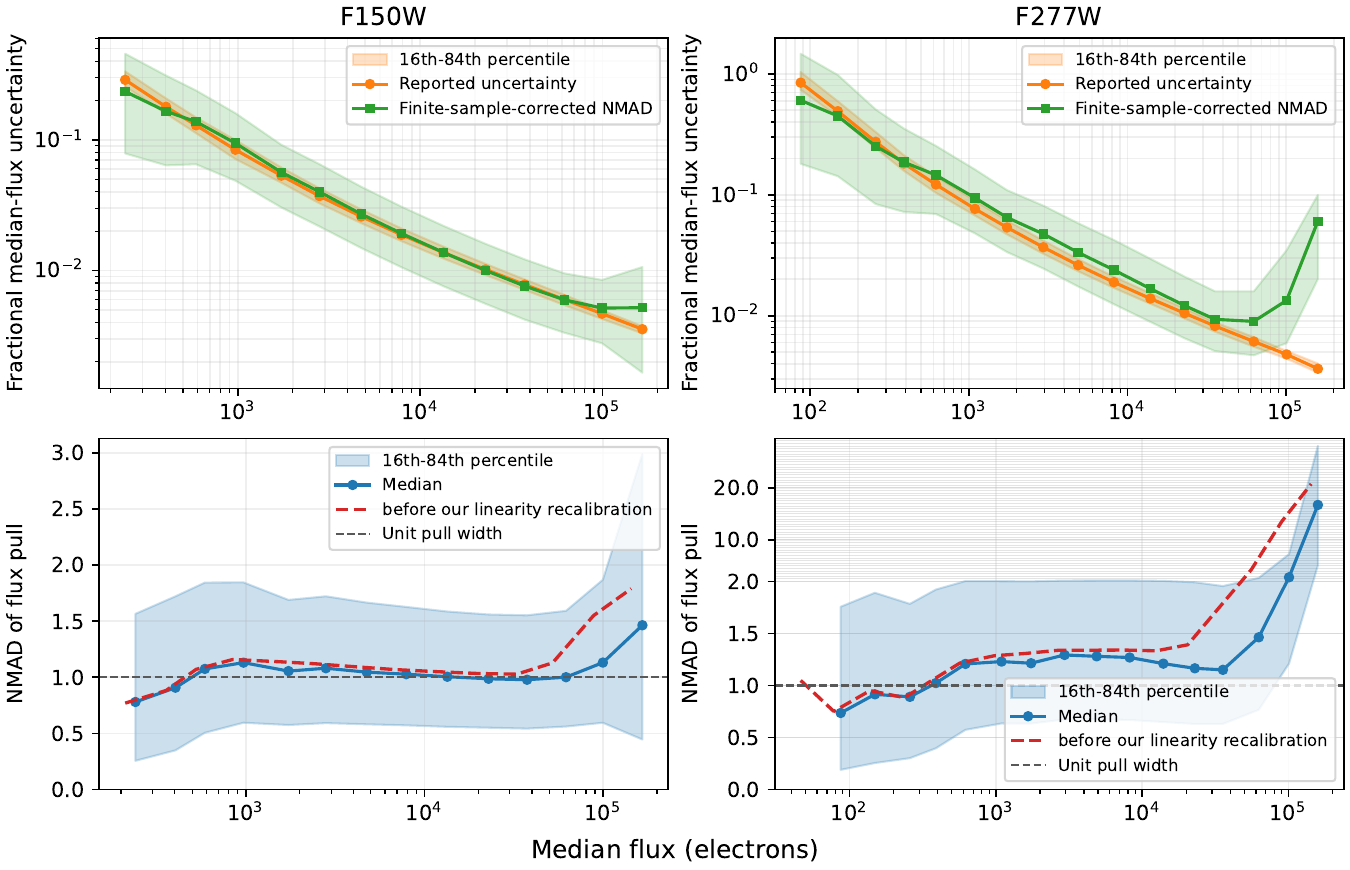}

    \caption{Performance of FLASH as a function of per-frame counts (in $e^-$). The {\bf top panels} show the fractional flux uncertainty in bins of flux and the NMAD dispersion (median absolute deviation scaled to give the same median value as for a Gaussian with the same number of samples) in the same bins. These curves track each other well until $\sim 10^5 e^-$ per frame. The {\bf bottom panels} show the NMAD pull in flux bins. These values are generally close to 1, indicating that the uncertainties are fairly accurate over most of the scientifically useful flux range. We also show the same curve generated with the photometry before our linearity recalibration (red dashed line). This comparison shows improvement for all flux levels, especially at high fluxes crucial to precise photometric measurements. The uncertainties are somewhat overestimated at low fluxes, indicating that the read noise may be overestimated. \label{fig:photperformance}}
\end{figure*}

Figure~\ref{fig:photperformance} shows the performance of FLASH as a function of flux. At fluxes of $\sim$~few $10^4~e^-$ where the noise floor $\lesssim 1\%$, our linearity calibration makes a large improvement in photometric precision.

\subsection{Fitting Radii} \label{sec:radii}

\newcommand{\radiussentence}{The modal star has a stellar radius $\sim 10\Rsol$.\xspace}

To assess our efficiency in finding microlensing events, we need an estimate of each star's radius. For this, we fit multi-band photometric measurements with four parameters per star: \Teff, \logg, radius, and extinction ($A_V$, assuming a \citealt{Fitzpatrick1999} extinction law with $R_V=3.1$). We compute our model synthetic photometry from the BOSZ library \citep{Bohlin2017, Meszaros2024}. Because we do not have spectroscopy, we fix [M/H], [$\alpha$/M], and [C/M] to 0 when computing our photometric grid. We use downhill-simplex \citep{NelderMead} for the optimization for each star.

\begin{figure}[htbp]
    \centering
    \includegraphics[width=0.5\linewidth]{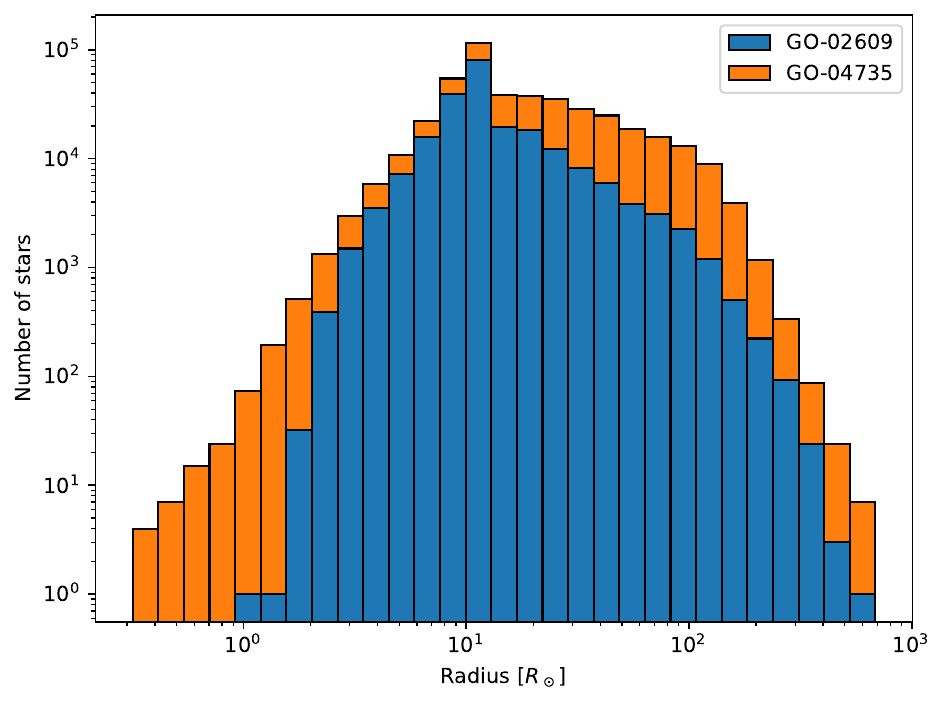}
    \caption{Fitted radii for the stars in the observations we examine. We cut a small number of failed fits with $\chi^2 > 200$. \radiussentence}
    \label{fig:radii}
\end{figure}

As it is helpful to have much bluer data to help constrain the tradeoff of extinction vs. \Teff, we merge the median flux of each star that we measure in F150W and F277W with the measurements from the Panchromatic Hubble Andromeda Treasury catalog (PHAT, \citealt{Williams2014PHATCatalog}), which adds F275W, F336W, F475W, F814W, F110W, and F160W (although not all bands are present for each star). Figure~\ref{fig:radii} shows a histogram for 442,127 stellar radii. \radiussentence

\section{Candidates}
\label{sec:candidates}

For our pair of filters (F150W and F277W) we fit light-curve templates to our light curves. We fit for four microlensing light-curve-fit parameters: the relative stellar radius/Einstein radius size ($\rho$ in Equation~\ref{eq:rho}), the time of closest approach, the relative speed on the sky of the lens, and the impact parameter. 

We compute a grid of model light curves, using \texttt{Mathematica} to numerically integrate the point-source lensing amplification (Equation~\ref{eq:lensingamp})
over a disk, assuming uniform surface brightness for the stars (limb darkening is small in the NIR, \citealt{Neilson2011}). We ignore wave optics, which we discuss more in Appendix~\ref{sec:wave}. We store and interpolate this grid for the light-curve fits. In principle, we should integrate over each \frametime exposure because some events will be undersampled at this cadence, but for this first selection stage, we do not integrate to reduce computation time.

For each star, we compute an inverse-variance-weighted mean for each band (ignoring bad photometry, see Section~\ref{sec:PSFPhotometry}) to compute a constant-flux $\chi^2$. Then, we repeatedly fit lensing light curves (to the data for both bands), running the minimization once for every observed time step, using that time as the initial event centroid. We use the downhill-simplex method \citep{NelderMead} to optimize without computing derivatives.

We select possible candidates that meet the following criteria; these criteria are not optimized, they are simply for selecting plausible candidates for our proof of concept. A future competitive program should optimize their cuts using simulated candidate injection.

\begin{itemize}
    \item The $\chi^2$ of the fit is lower than the $\chi^2$ of a constant flux by 24.5021 (two-tailed $4\sigma$ for four additional degrees of freedom compared to the constant light-curve fit).
    \item After obtaining the best fit, we separately scale the template in F150W and F277W and obtain separate amplifications. We require the inferred F150W and F277W amplifications to
    agree within $2.5\sigma$:
    \begin{equation}
            \frac{
        \left|a_{\mathrm{F150W}}-a_{\mathrm{F277W}}\right|
    }{
        \sqrt{
            \sigma^2_{a,\mathrm{F150W}}
            +
            \sigma^2_{a,\mathrm{F277W}}
        }
    }
    < 2.5.
    \end{equation}
    \item To ensure that the data are valid over the light curve, we define the relevant time range as the observations for which
    \begin{equation}
    A(t)-1 >
    0.05\left[A_{\max}-1\right].
    \end{equation}
    Within this range, we require valid measurements for more than $75\%$
    of the measurements in each band.
\end{itemize}
For each distinct event passing these cuts, we write out a light-curve plot to disk (sometimes, more than one light curve is written for the observations of a single star because of local minima in the $\chi^2$; this is one reason why we initialize the light-curve fits with many initial event centroids). 
In all, we save 30,638 candidate plots for 11,695 stars.

For this paper, we chose a subset of events to examine by eye, meeting these criteria:
\begin{itemize}
    \item Normalized median absolute deviation (a robust dispersion) pulls (residuals/uncertainties) for the best fit are less than 1.4 in both the F150W and F277W, indicating a reasonable fit for most data points. After this cut, 10,081 candidate events remained. We checked the 2D images for a few randomly selected failed events, and the failures were due to unflagged bad pixels, so a dedicated competitive program should update the bad pixel list \citep{Currie2018}.
    \item In contrast to the above cut, which only required consistency between the F150W and F277W, we required S/N $> 4$ in both the F150W and F277W individually. One could consider relaxing this cut in future work, but we wanted the most plausible candidates for now. After this cut, 8,345 candidates remained.
    \item To help remove variable stars, which are generally variable on much longer timescales, we filtered out long-timescale events. We required that the lens crosses the source radius on the sky in less than 50 timesteps ($t_*<1074$ seconds). Given our short time baseline, we cannot really investigate longer timescales than this. After this cut, 5,114 candidates remained.
    \item Finally, we require that the inferred dynamic range during our observations was $> 5\%$. After this cut, 480 candidate events remained. 
\end{itemize}

\begin{figure*}[htbp]
    \centering
    \includegraphics[width=0.49\textwidth]{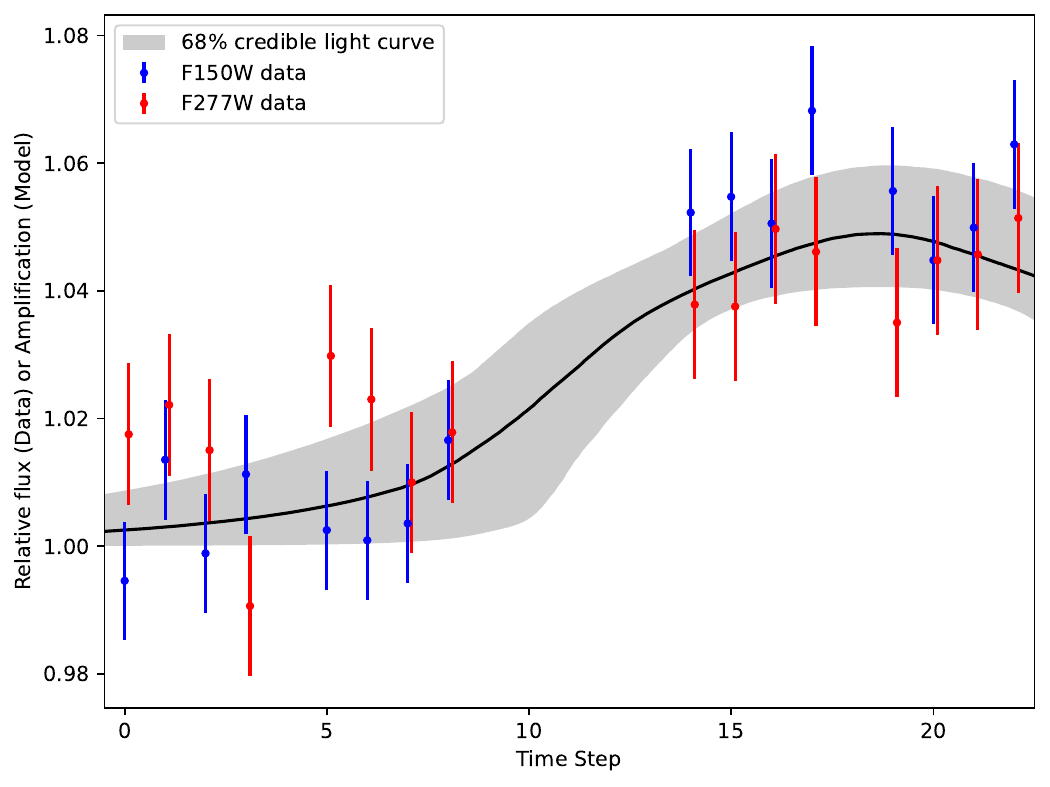}

    \includegraphics[width=0.49\textwidth]{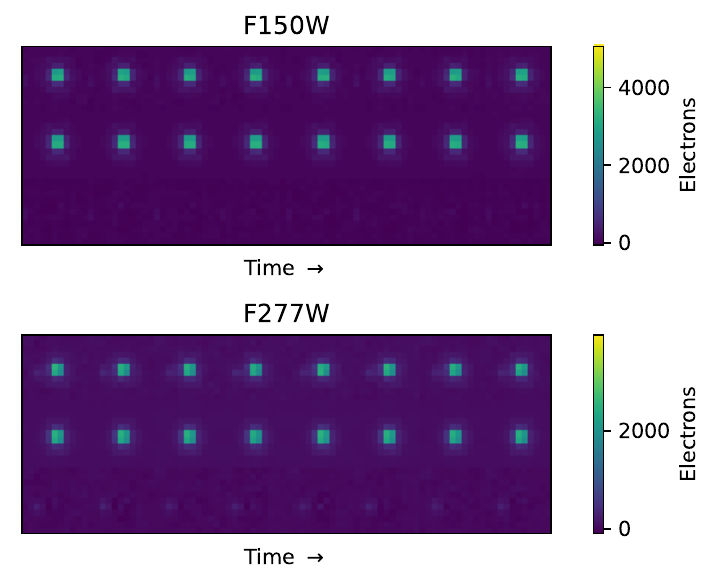}
    \includegraphics[width=0.49\textwidth]{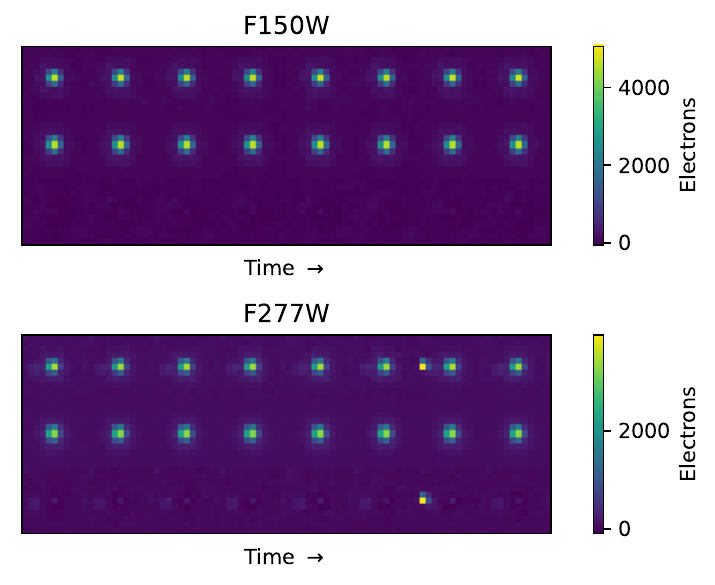}
    \caption{{\bf Top panel:} Light curve for star 139069 in both the F150W and F277W. The gray band is the 68\% credible interval of lensing amplification models. The data provide a baseline to constrain the unlensed flux, but it is still insufficient to establish microlensing.
    {\bf Lower panels:} 2D image cutouts for both the F150W ({\bf top}) and F277W ({\bf bottom}). We show the cutout, the PSF model, and the residuals from top to bottom for each filter. The residuals are fairly clean, but the F277W suggests another nearby star. Note the cosmic ray hit in the left F277W panel, which is nevertheless well corrected for during the flux determination. The changed subpixel position visually exaggerates the amplification between the two panel sets. This candidate clearly shows why we should avoid frequent dithering for microlensing measurements.}
    \label{fig:candtwo}
\end{figure*}

\begin{figure*}[htbp]
    \centering
    \includegraphics[width=0.49\textwidth]{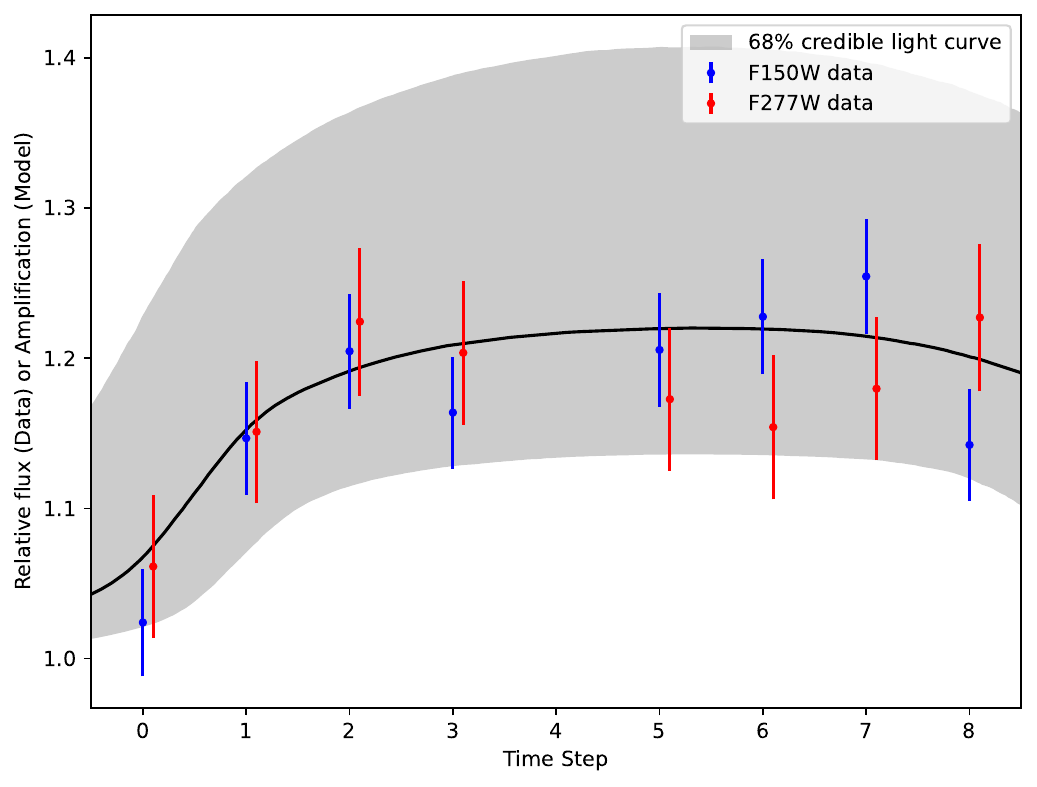}
    \includegraphics[width=0.49\textwidth]{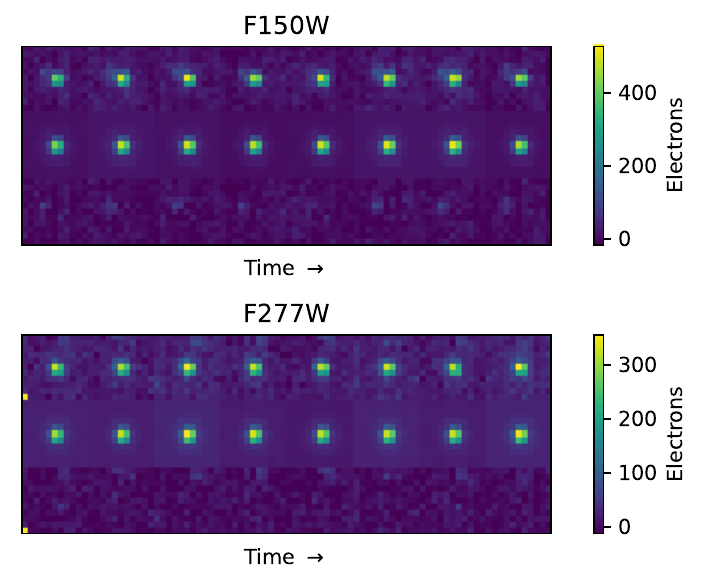}

    \caption{{\bf Left panel:} Light curve for star 213169 in both the F150W and F277W. The gray band is the 68\% credible interval of lensing amplification models. The data do not have sufficient time baseline to strongly constrain the unlensed flux, so amplifications from $\sim 13\%$ to $\sim 40\%$ are consistent with the data. {\bf Right panels:} 2D image cutouts for both the F150W ({\bf top}) and F277W ({\bf bottom}). We show the cutout, the PSF model, and the residuals from top to bottom for each filter. The residuals are fairly clean, but do show two nearby stars. The lower flux in the first F150W image is visible by eye. In all, this candidate has a plausible light curve for finite-size microlensing but the observations lack the time baseline to support this explanation.}
    \label{fig:candone}
\end{figure*}

We visually inspected all 480 candidate light-curve fits. Of these, we inspected several 2D image sequences. We found no clearly convincing candidates. Figures~\ref{fig:candtwo} and \ref{fig:candone} show two representative examples of the most credible. Figure~\ref{fig:candtwo} shows a jump in flux between two dithers that is statistically significant. These occurrences were fairly common, so a program seeking competitive constraints should avoid frequent dithering for better stability. Figure~\ref{fig:candone} shows an increase in flux in one dither position, so it is more credible. This type of event is much rarer, but it is difficult to believe it is microlensing without a much longer time baseline.

We thus compute a 95\% upper limit assuming zero events. Assuming Poisson statistics, this limit is $-\log(0.05) \approx 2.996/$expected number of events.

\section{Primordial Black Hole Constraints} \label{sec:PBHConstraints}

We perform a Monte-Carlo simulation to evaluate our upper limits on the primordial black hole density. The details are in Appendix~\ref{app:event_rate_mc}, but we summarize here. For each star, for each filter, we take the estimated photometric precision, the duration of our observations (measured in Section~\ref{sec:data}), and the estimate of the stellar radius (Section~\ref{sec:radii}) and place the star into one of many bins with similar stars, keeping track of the number of stars per bin.

For each star, we place an appropriate number of black holes, each with a randomly chosen velocity perpendicular to the line of sight and location. We sample from the DM halos of the Milky Way and M31, assuming \cite{Navarro1997} dark-matter profiles.

We compute light curves over the same timescale each star was observed over and flag events that change by 4~$\sigma$ (but now per point $\Delta \chi^2=16$). This approximates the procedure we used to flag and inspect candidate events. In principle, we should use the exact same fitting procedure (Section~\ref{sec:candidates}), but as this work is a proof of concept, we leave this computationally intensive step for future work.

\newcommand{\massradiussentence}{We see that our constraints are driven by stars in the $\sim$ 5--100~$R_{\odot}$ range. We also see that these observations are not long enough to have a meaningful contribution from lenses in M31.\xspace}

\begin{figure*}
\centering
    \includegraphics[width = 0.8\textwidth]{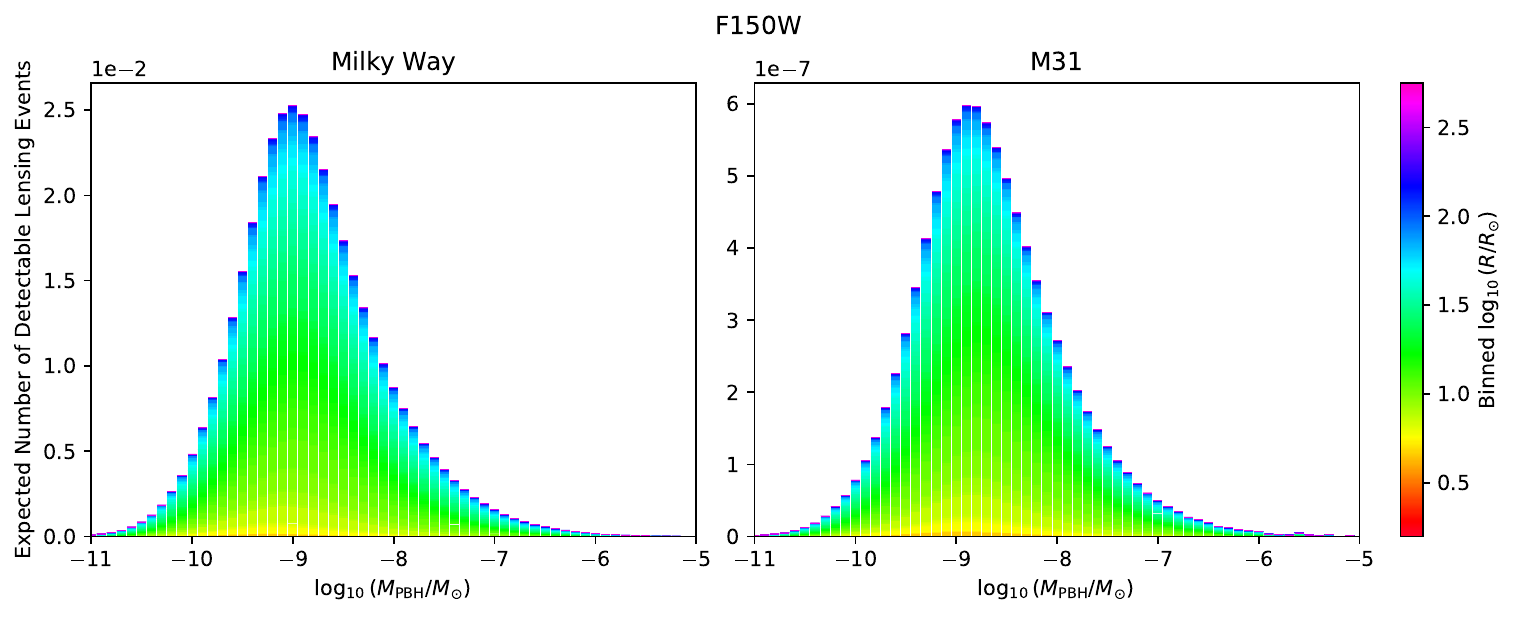}

    \caption{Expected number of events as a function of PBH mass from our MC simulation assuming $\fPBH = 1$ for lenses in the Milky Way ({\bf left panel}) and M31 ({\bf right panel}). The number of events is (significantly) less than 1, so we do not expect any real events in a sample of this size. To help optimize future constraints, we plot a histogram of our MC trial masses and color-code by which radius stars contribute to the constraints. \massradiussentence \label{fig:EventsByMassRadius}}
\end{figure*}

Figure~\ref{fig:EventsByMassRadius} shows our Monte Carlo calculation, binned by stellar radius for a range of assumed primordial black hole masses. Even in the most sensitive mass range of $\sim 10^{-9} \Msol$, we only have a $\sim 3\%$ chance of observing an event in our dataset assuming all dark matter is primordial black holes. \massradiussentence

As with other microlensing surveys, our sensitivity has a unimodal peak in PBH mass. For large masses, the number of events drops (for a fixed PBH density), and the timescales of the events increase according to Equation~\ref{eq:tE} and extend beyond the window these programs monitored. For small PBH masses, there are many events, but they are extremely weak due to finite source effects and (for even smaller masses, wave optics). The transition to the finite-source regime occurs when the lens-plane Einstein radius becomes comparable to the stellar radius projected to the lens plane, $R_E \simeq xR_\star$.  Equivalently, the Einstein angle is comparable to the angular stellar radius, $\theta_E\simeq\theta_\star$.  Solving this condition for the lens mass gives a characteristic finite-source transition mass
\begin{equation}
M_{\rm fs} =
\frac{c^2}{4G}
\frac{xR_\star^2}{D_s(1-x)}
\simeq
3.5\times10^{-11}M_\odot
\left(\frac{x}{0.01}\right)
\left(\frac{1-x}{0.99}\right)^{-1}
\left(\frac{R_\star}{R_\odot}\right)^2
\left(\frac{D_s}{760~{\rm kpc}}\right)^{-1}.
\end{equation}
For $M\gtrsim M_{\rm fs}$, the event timescale and peak amplification are primarily controlled by the Einstein radius.  For $M\lesssim M_{\rm fs}$, the source is resolved by the lensing geometry: the peak amplification is suppressed, and the observable duration approaches the projected stellar-radius crossing time rather than continuing to decrease as $M^{1/2}$.

\newcommand{\tenthousandsentence}{To show the constraint on the same scale as literature constraints, we show our constraint 10,000$\times$ stronger than what we could obtain from this (suboptimal for PBH constraints) dataset. This 10,000$\times$ is not a forecast, but Section~\ref{sec:conclusions} sketches how this order of magnitude is likely to be obtainable with an optimized survey design.\xspace}

\begin{figure*}
\centering
\includegraphics[width=0.8\textwidth]{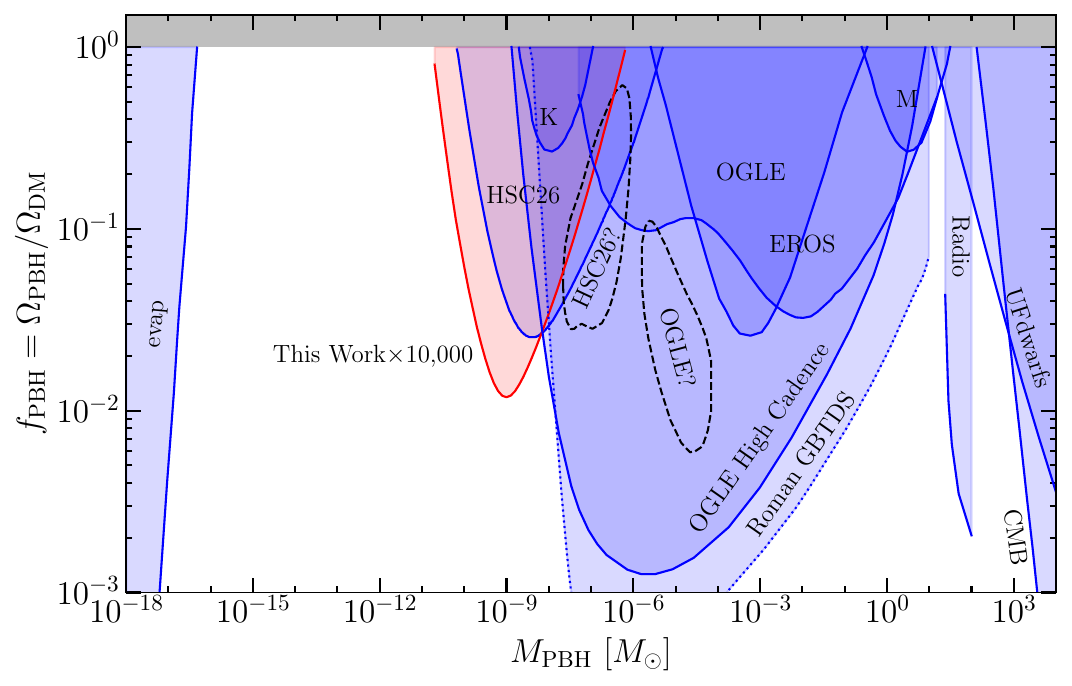}

\caption{Our 95\% upper limits on the fraction of dark matter that can be primordial black holes as a function of mass, plotted with the \cite{Kavanagh_PBHbounds_2019} package. \tenthousandsentence  \label{fig:massconstraints}}
\end{figure*}

Figure~\ref{fig:massconstraints} compares our constraints to constraints from the literature using the \cite{Kavanagh_PBHbounds_2019} package.\footnote{We include: PBH evaporation \citep{Carr2010}, HSC \citep{Sugiyama2026HSCM31PBH}, Kepler \citep{Griest2014}, OGLE \citep{Mroz_2024a,Mroz_2024b}, OGLE High Cadence \citep{Mroz_2024highCadence}, OGLE? \citep{Niikura2019bPhysRevD}, EROS \citep{Tisserand2007EROS2}, the Roman Space Telescope GBTDS forecast \citep{DeRocco2024}, MACHO \citep{Alcock2001}, Radio \citep{Manshanden2019}, CMB \citep{Ali-Haimoud2017}, and Ultra Faint Dwarfs \citep{Brandt2016}.} At our most sensitive PBH mass, we expect $\sim 0.03$ events for $\fPBH=1$, so our 95\% upper limit is $\fPBH < 100$ in this mass range. \tenthousandsentence

\section{Conclusions and Discussion}
\label{sec:conclusions}

This work presents \textsc{FLASH}: Fast Lensing And Sub-minute High-accuracy photometry with JWST, a pipeline designed to recover rapid microlensing events (more precisely, nanolensing) with JWST/NIRCam imaging by preserving detector-ramp time information and performing high-precision crowded-field photometry. We demonstrated the pipeline's effectiveness on two archival JWST programs. These programs were suboptimal for microlensing; nevertheless, they allowed us to estimate PBH sensitivity. Our constraints are based on radius fitting from multi-band \textit{HST}/JWST observations and then Monte Carlo simulating lensing events from a galactic/M31 DM model appropriate for \emph{each} star. This procedure takes into account correlations between magnitudes, color, and radii.

In the end, we obtain a constraint that $\fPBH < 10^2$ for PBH masses $\sim 10^{-8.5} \Msol$ at 95\% confidence from $\sim 1$ hour of F150W/F277W observations.

An optimal JWST observing strategy would improve on these constraints by orders of magnitude. 1) Using the F150W2 and F322W2 would give $4\times$ and $2\times$ the signal per second and thus up to $\sim$~2--4$\times$ the signal-to-noise, as our observations are limited by read noise at the faint end (and the W2 filters would deliver many times more stars at the same S/N). 2) A better M31 field choice would enable $\sim 10$--100$\times$ the stars as the fields we looked at (as judged from the PHAT and PHAST programs, \citealt{PHAT, PHAST}). 3) Eliminating dithering would improve stability (perhaps a compromise would be a small dither every few hours to improve the archival value of the data). 4) One could expose for much longer, which would deliver both more chances for events to happen and better establish the unlensed-flux baseline for any events that are seen. In addition, a longer time baseline would open up discovery space for M31-hosted lenses, which have an enormous lensing cross section, but produce much slower events. 5) The NIRCam readout strategy could be optimized in concert with optimizing the choice of field to match the timescale of the expected lensing events. This work is thus a successful proof of concept that demonstrates the promise of a dedicated program.

Beyond obtaining optimized JWST data, several future pipeline improvements will make it more efficient. We found that a few spot-checked candidates with large NMAD residuals were contaminated by unflagged bad pixels. This finding motivates improved bad-pixel flagging for NIRCam (e.g., using the method of \citealt{Currie2018}). When a more competitive dataset becomes available, we can verify the accuracy of our constraints with end-to-end (pixels-to-lightcurves) Monte Carlo injections of simulated events. This would also help to optimize some of the selection cuts we made on the events we examined. With improved pixel flagging, we could even extend the pipeline into the pixel lensing regime \citep{Gould1996PixelLensing}.

In addition to lensing, other potential applications of FLASH include searches for rapid stellar variability that is diluted or lost in standard integration-level products such as stellar flares, pulsations, and serendipitous stellar occultations by Solar System bodies (when targeting observations, faster subarrays are more optimal). Applied broadly to archival NIRCam observations, it could turn many ordinary imaging programs into a large survey for fast infrared transients and variable stars.

\begin{acknowledgments}
The technical support and advanced computing resources from University of Hawaii Information Technology Services – Research Cyberinfrastructure, funded in part by the National Science Foundation CC* awards \#2201428 and \#2232862 are gratefully acknowledged. IS acknowledges NASA ROSES grants 80NSSC24K1489 and 24-ADAP24-0074, and contract number 80NM0018F0610 via a JPL sub-award. DR acknowledges support for this work from the ChatGPT for Academic Researchers program. This work is based [in part] on observations made with the NASA/ESA/CSA James Webb Space Telescope. The data were obtained from the Mikulski Archive for Space Telescopes at the Space Telescope Science Institute, which is operated by the Association of Universities for Research in Astronomy, Inc., under NASA contract NAS5-03127 for JWST. These observations are associated with programs GO-4735 and GO-2609.
\end{acknowledgments}

\section{AI-use disclosure}

Generative AI tools were used in the preparation of this work as follows: ChatGPT 5.2--5.6 (06/2025-07/2026) for pipeline code generation, debugging, code verification, drafting, and grammar editing (but not other writing). No confidential collaboration data were uploaded to unapproved systems.  
The authors take full responsibility for the scientific claims, code, figures, and interpretations.

\facilities{JWST(NIRCam)}

\software{Astropy \citep{2013A&A...558A..33A,2018AJ....156..123A,2022ApJ...935..167A},
JWST Pipeline \citep{JWSTpipeline},
Kavanagh \citep{Kavanagh_PBHbounds_2019},
Matplotlib \citep{Matplotlib},
NumPy \citep{NumPy},
Mathematica \citep{Mathematica},
PhotUtils \citep{photutils}}

\appendix

\section{Wave-Optics Suppression} 
\label{sec:wave}

\begin{figure*}[htbp]
\centering
   \includegraphics[width=0.47\textwidth]{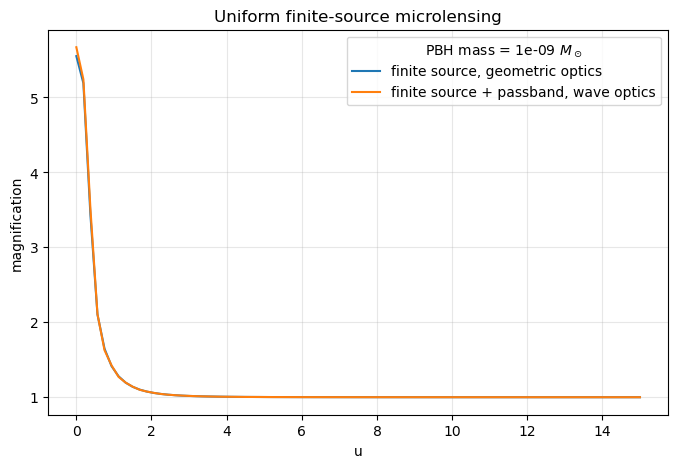}
   \includegraphics[width=0.47\textwidth]{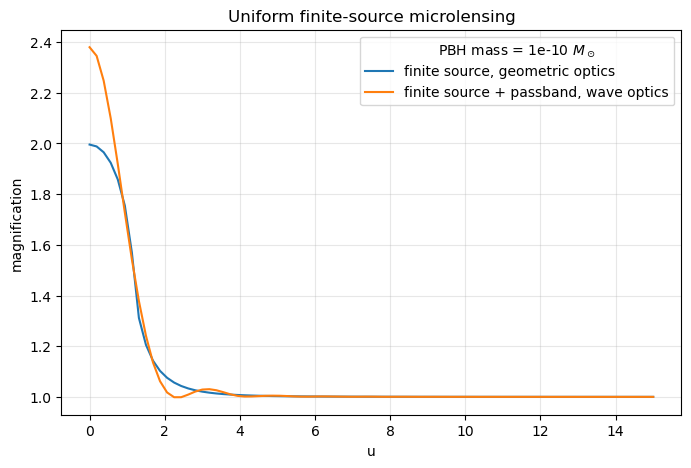}
    \includegraphics[width=0.47\textwidth]{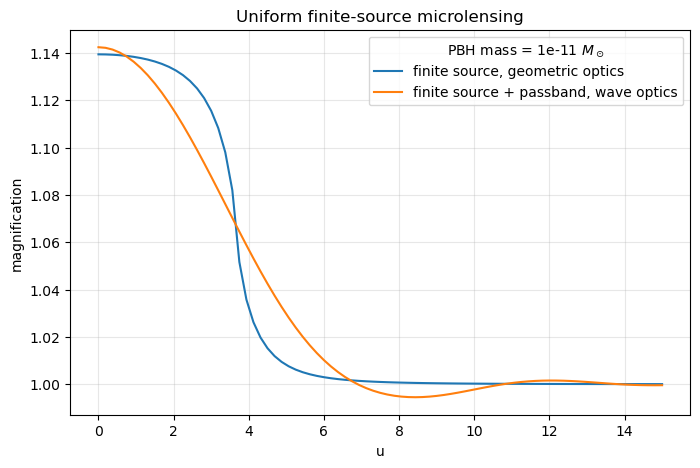}
    \includegraphics[width=0.47\textwidth]{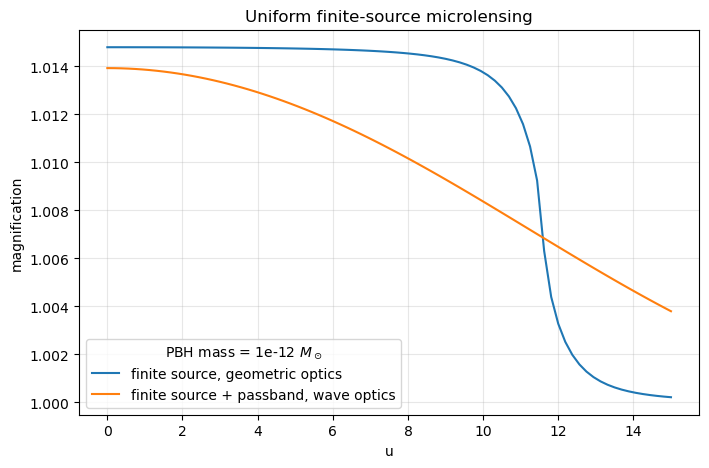}
    \caption{Comparison of microlensing amplifications calculated from geometric and wave optics for an $M = 10^{-12}M_\odot - 10^{-9}M_\odot$ PBH as a function of impact parameter. The wavelength band is assumed to be $2.4\mu-4\mu$ ($\sim$~F322W2, which we recommend for the red channel of NIRCam for dedicated work) for a $2R_\odot$ star in M31, lensed by a lens at $8$kpc. The figure shows that wave-optics corrections can be safely neglected for our purposes, although they might be useful for extremely low PBH masses. \label{fig:waveoptics}}
\end{figure*}

Wave optics can erase the lensing effect if the wavelength of light is close to the Schwarzschild radius \citep{Schneider1992}. At sufficiently low lens mass and long wavelength, diffraction suppresses the geometric-optics magnification.  A standard dimensionless wave-optics parameter is
\begin{equation}
w \equiv \frac{8\pi GM}{c^2\lambda}.
\label{eq:wparam}
\end{equation}
For $w\gg 1$, the geometric-optics limit is recovered; for $w\ll 1$, lensing is strongly suppressed and becomes chromatic.  The transition $w\simeq 1$ occurs at
\begin{equation}
M_w \simeq
5.4\times10^{-11}M_\odot
\left(\frac{\lambda}{2~\mu{\rm m}}\right).
\label{eq:mw}
\end{equation}
Thus, in the near-infrared, wave optics becomes important at masses similar to the finite-source transition for solar-radius stars in M31.  This does not eliminate the usefulness of JWST, but it means that forecasts below a few $\times10^{-11}M_\odot$ must include wavelength-dependent finite-source wave-optics templates. In Figure~\ref{fig:waveoptics} we show examples of calculated wave optics amplifications to illustrate that they are not important even for long-wavelength F322W over the mass range to which the present archival data have appreciable sensitivity, but they become important for a future JWST experiment attempting to extend sensitivity below $10^{-11}M_\odot$.

\section{Monte Carlo Event-Rate Estimate as a function of precision and radius}
\label{app:event_rate_mc}

Since the present data are suboptimal for lensing, we do not perform a
pixel-level full injection--recovery calculation to refine our derived
constraints. Instead, we combine an analytic event-rate calculation with a
Monte Carlo estimate of the efficiency for the observing-window. We
evaluate these quantities as a function of stellar radius and photometric
precision, using the stellar radius assigned to each catalog star from the
JWST photometry combined with PHAT-based stellar-population information and
our estimate of its photometric uncertainty. This approach captures the
dominant physical limitation in the low-mass regime, finite-source
suppression, while retaining the correlations among stellar radius, color,
magnitude, and photometric precision. This method is therefore preferable to assuming a single overall stellar-radius distribution.

For each source star $j$, with source distance $D_s$ and estimated stellar
radius $R_{\star,j}$, we consider lenses at distance $D_l=s$ along the M31
line of sight and define $x=s/D_s$. The Einstein radius in the lens plane is
\begin{equation}
R_E(M,s) =
\left[
\frac{4GM}{c^2}D_s x(1-x)
\right]^{1/2},
\end{equation}
and the finite-source parameter is
\begin{equation} \label{eq:rho}
\rho_j(M,s) =
\frac{\theta_{\star,j}}{\theta_E}
=
\frac{xR_{\star,j}}{R_E(M,s)}.
\end{equation}

The code computes a finite-source magnification threshold rather than
assuming a point source. For a uniform stellar disk, the finite-source
magnification is obtained by averaging the point-source magnification over
the projected stellar disk,
\begin{equation}
A_{\rm fs}(u,\rho)
=
\frac{1}{\pi\rho^2}
\int_{|\boldsymbol{r}|<\rho}
A_{\rm ps}\left(|\boldsymbol{u}-\boldsymbol{r}|\right)
\,d^2\boldsymbol{r},
\end{equation}
where
\begin{equation} \label{eq:lensingamp}
A_{\rm ps}(u)=
\frac{u^2+2}{u\sqrt{u^2+4}}.
\end{equation}
For a chosen threshold $A_{\rm th}$, the threshold impact parameter
$u_T(\rho;A_{\rm th})$ is defined by
\begin{equation}
A_{\rm fs}\left[u_T(\rho),\rho\right]=A_{\rm th}.
\end{equation}
If the maximum finite-source magnification, $A_{\rm fs}(0,\rho)$, is below
$A_{\rm th}$, the threshold is set to $u_T=0$. In practice, the code
precomputes $u_T$ on a logarithmic grid in $\rho$ and interpolates this table
for each source and lens distance. This makes the calculation fast enough to
run over many catalog stars and PBH masses.

The analytic rate calculation uses a line-of-sight halo density model to
compute the optical depth and event rate for each star. In the implementation
used here, the halo density $\rho_{\rm halo}(s)$ is an NFW profile
\citep{Navarro1996,Navarro1997}. For the Milky Way halo, we adopt a scale
radius $r_s=16\,{\rm kpc}$ and normalize the profile to the local dark-matter
density $\rho_{\rm local}=0.4\,{\rm GeV\,cm^{-3}}$. The line of sight is set
to the direction of M31, $l=121.17^\circ$, $b=-21.57^\circ$. For the M31
halo, we use $r_s=25\,{\rm kpc}$ and an M31 distance of
$D=760\,{\rm kpc}$. The Milky Way and M31 contributions are calculated
separately, using their respective halo and velocity distributions, and are
summed at the end. For the present data set, the M31 contribution is
negligible because the observing windows are too short to efficiently sample
the longer-duration events produced by most M31 lenses, but it can become
important for longer observations.

The PBH number density is
\begin{equation}
n_{\rm PBH}(s) =
\frac{f_{\rm PBH}\rho_{\rm halo}(s)}{M},
\end{equation}
where $f_{\rm PBH}$ is the fraction of dark matter in PBHs of mass $M$. The
finite-source optical depth for star $j$ is then
\begin{equation}
\tau_j(M) =
\int_0^{D_s}
n_{\rm PBH}(s)\,
\pi\left[u_T(\rho_j)R_E(M,s)\right]^2
\,ds,
\end{equation}
and the corresponding event rate is
\begin{equation}
\Gamma_j(M) =
\int_0^{D_s}
n_{\rm PBH}(s)\,
2u_T(\rho_j)R_E(M,s)\,
\bar{v}_\perp
\,ds,
\end{equation}
where $\bar{v}_\perp$ is the mean transverse speed for the adopted velocity
distribution. This expression makes explicit that finite-source effects
enter both through the geometric cross section and through the elimination of
source--lens geometries for which $A_{\rm fs}(0,\rho)<A_{\rm th}$.

The Monte Carlo portion of the calculation uses the observing-window
and photometric precision for each source. Rather than drawing lens distances
directly from $\rho_{\rm halo}(s)$, we importance-sample them from the
spatial part of the differential event-rate distribution. For fixed lens
mass and source distance,
\begin{equation}
\frac{d\Gamma}{ds}
\propto
\rho_{\rm halo}(s)\,
u_T(\rho_j)R_E(M,s),
\end{equation}
apart from factors independent of $s$. Since
$R_E\propto [x(1-x)]^{1/2}$, the lens-distance proposal distribution used by
the Monte Carlo is
\begin{equation}
q(s)
\propto
\rho_{\rm halo}(s)\,
u_T(\rho_j)
\left[x(1-x)\right]^{1/2}.
\label{eq:sproposal}
\end{equation}
This sampling efficiently concentrates trials in lens geometries that
actually contribute to the event rate and avoids spending trials in regions
where finite-source suppression gives $u_T=0$.

For each trial event, a transverse velocity is drawn from the underlying
two-dimensional Gaussian velocity distribution, so that the magnitude
$v_\perp$ follows the corresponding Rayleigh distribution. The code then
computes $R_E$, $\rho$, $u_T$, and the Einstein time,
\begin{equation}
t_E =
\frac{R_E}{v_\perp}.
\end{equation}
Because the differential event rate contains an additional factor of
$v_\perp$, the Monte Carlo trials are subsequently weighted by $v_\perp$
when computing the efficiency.

For feasible events with $u_T>0$, the closest-approach impact parameter is
drawn uniformly across the lensing tube,
\begin{equation}
p(u_0)\,du_0
=
\frac{du_0}{u_T},
\qquad
0<u_0<u_T.
\label{eq:u0distribution}
\end{equation}
Note that this differs from the $p(u)\propto u$ area weighting appropriate to an
instantaneous optical-depth calculation. For an event rate, trajectories
crossing a strip of width $2u_TR_E$ are uniform in their closest-approach
distance $u_0$.

For each trial, the half-duration for which the instantaneous magnification
exceeds the absolute threshold $A_{\rm th}$ is
\begin{equation}
\Delta t_{\rm th,half}
=
t_E
\left(u_T^2-u_0^2\right)^{1/2},
\end{equation}
so that the full above-threshold duration is
$2\Delta t_{\rm th,half}$. A random time of closest approach $t_0$ is drawn
uniformly over the observed time baseline. As a computational preselection,
the event is retained only if at least one observation in the source's
actual time array falls within
\begin{equation}
|t_i-t_0| < \Delta t_{\rm th,half}.
\end{equation}

For the surviving trials, we then evaluate the finite-source magnification
at the sampled epoch nearest to $t_0$, which we denote $A_{\rm near}$, and
at the sampled epoch farthest from $t_0$, denoted $A_{\rm far}$. Because
these relatively short observing sequences do not necessarily contain an
independently established unlensed baseline, we use the observed dynamic
range rather than the absolute peak magnification as the detection
criterion. A trial is counted as detected when
\begin{equation}
\frac{A_{\rm near}}{A_{\rm far}}
\geq A_{\rm th}.
\label{eq:dynamicrange}
\end{equation}
For the calculations presented here, $A_{\rm th}$ is set from the measured
fractional photometric uncertainty of the relevant light curve; for example,
a nominal $4\sigma$ threshold corresponds to
$A_{\rm th}=1+4\sigma_{\rm phot}$. This criterion is intended to approximate
the event-selection procedure used for the present proof-of-concept data
rather than a full end-to-end injection--recovery analysis.

The spatial proposal distribution in Equation~(\ref{eq:sproposal}) already
contains the factors
$\rho_{\rm halo}(s)u_T R_E$ appearing in the event rate. The remaining
velocity dependence is included by weighting each Monte Carlo trial by
$v_\perp$. The event-rate-weighted detection efficiency for source $j$ is
therefore
\begin{equation}
\epsilon_j(M)
=
\frac{
\displaystyle
\sum_{k\in{\rm detected}} v_{\perp,k}
}{
\displaystyle
\sum_{k\in{\rm trials}} v_{\perp,k}
}.
\label{eq:mceff}
\end{equation}
Here the sums extend over trials drawn from the spatial proposal distribution
of Equation~(\ref{eq:sproposal}); geometries with $u_T=0$ carry no event-rate
weight. The Monte Carlo uncertainty in $\epsilon_j$ is estimated from the
weighted trial distribution, including the reduction in effective sample
size associated with the velocity importance weights.

The expected number of detected events is obtained by combining the analytic
event rate with the Monte Carlo efficiency for every source:
\begin{equation}
\Lambda(M) =
\sum_j
\Gamma_j(M)
T_j
\epsilon_j(M),
\end{equation}
where $T_j$ is the time baseline of the light curve for source $j$. The
probability of detecting at least one event is then
\begin{equation}
P(\geq 1) =
1-\exp[-\Lambda(M)].
\end{equation}
The calculation is repeated over PBH masses to estimate the scaling of JWST
sensitivity with $M$. We tested the analytic portion of our implementation
against {\tt LensCalcPy}\footnote{https://github.com/NolanSmyth/LensCalcPy}
for several special cases and found agreement at the subpercent level for
both halos.

{}

\end{document}